\documentclass[reprint,amssymb, amsmath, aip,cha,twocolumn]{revtex4-1}
\usepackage{graphicx}
\usepackage[caption = false]{subfig}
\usepackage{color}
\usepackage{epsfig}
\usepackage{ifpdf}
\usepackage{url}%
\usepackage{cleveref}
\usepackage{bm}
\usepackage[colorlinks=true,linkcolor=blue]{hyperref}%
\expandafter\ifx\csname package@font\endcsname\relax\else
 \expandafter\expandafter
 \expandafter\usepackage
 \expandafter\expandafter
 \expandafter{\csname package@font\endcsname}%
\fi
\begin{document}
\title{Effect of size and shape of a moving charged object on the propagation characteristics of precursor solitons}%
\author{Garima Arora}%
\email{garimagarora@gmail.com}
\author{P. Bandyopadhyay}
\author{M. G. Hariprasad}
\author{A. Sen}
\affiliation{Institute For Plasma Research, HBNI, Bhat, Gandhinagar, Gujarat, India, 382428}%
\date{\today}
\begin{abstract} 
We report on experimental observations on the modifications in the propagation characteristics of precursor solitons due to the different shapes and sizes of the object over which the dust fluid flows. The experiments have been performed in a $\Pi$ shaped Dusty Plasma Experimental (DPEx) device where dusty plasma is created in a DC glow discharge Ar plasma using kaolin particles. A floating copper wire installed radially on the cathode, acts as a charged object in the plasma environment. The flow on the dust fluid is initiated by suddenly lowering the potential of the charged object from grounded potential to close to floating potential. The size (height and width) of the potential hill is then varied by drawing current from the wire through a variable resistance. 
With a decrease in the height of the potential hill the amplitude, velocity and the number of excited precursor solitons are found to decrease whereas the widths of the solitons are seen to increase. It is found that below a threshold value these solitary waves are not excited and the dust fluid simply flows over the hill. To examine the effect due to the shape of the potential profiles, the wire is replaced by a triangular object. Only trailing wakes are seen to be excited when the dust fluid faces the linearly increasing slope of the potential profile whereas both solitons and wakes get excited when the object is placed with the sharp edge facing the flow. All the experimental findings qualitatively agree with numerical solutions obtained with different source terms in the forced-Korteweg de Vries (f-KdV) model equation.
\end{abstract}
\maketitle
\section{Introduction}\label{sec:intro}
 The simultaneous excitation of precursor solitons ahead of a fast moving object with wakes behind it, is a phenomenon  that has been widely studied in hydrodynamics particularly in the context of disturbances created by ships and boats close to the coast \cite{huang1982ships,ertekin,ertekin1986waves}. These earlier studies have also provided model descriptions of this fascinating non-linear phenomenon by employing a forced Korteweg de Vries equation or a forced generalized Boussinesque equation \cite{wu1987generation}. Controlled laboratory experiments using model ships moving in a channel were carried out by Huang \textit{et al.} \cite{huang2010dynamics} and Sun \cite{sun1985evolution}. It was found that as long as the speed of the object moving through the fluid (water) was below a critical value (in this case the phase velocity of the surface water waves) the movement only created the customary trailing waves (wakes) in the downstream direction. A dramatic change occurred when the object speed was trans-critical. Then, in addition to the trailing wakes the object also created a steady stream of solitons ahead of it in the upstream direction. These solitons with a speed higher than the object moved away as precursors. The phenomenon was Galilean invariant \cite{wu1987generation} - it could be reproduced by keeping the object stationary and moving the fluid over it.\par

Fluid concepts have often been successfully translated and exploited to understand collective phenomena in plasmas.  
A prime example is the extensive past study of linear waves and nonlinear structures in a dusty (or complex) plasma medium. Such a medium, consisting of heavy dust grains immersed in a plasma of electrons and ions, display a host of collective excitations on a slow time scale due to the low charge to mass ratio of the dust particles and can be understood from a fluid model of the dusty plasma. There is a rich literature on the study of linear waves \textit{e.g.} the Dust Acoustic Wave (DAW) \cite{shukladaw,barkandaw}, the Dust Ion Acoustic Wave (DIW)\cite{shukladiaw}, Dust Lattice Waves (DL)\cite{franklattice} and non-linear waves like Dust Acoustic Solitary Waves (DASw)\cite{rao1994nonlinear,pintudasw}, Shock Waves \cite{theoretical_shock,heinrichshock}, dust voids \cite{thomas_void,goree_void,Khrapak_void}, Vortices \cite{manjit_vortex,morfill_wake,mangilal_vortex} etc \cite{usachev2004project} consisting of both theoretical and experimental investigations.
As in a hydrodynamic fluid, there have also been observations of wake structures in laboratory dusty plasmas \cite{morfill_wake, Dubin_wake,Hou_wake}.  These wakes (sometimes termed as Mach cones \cite{Samsonov_mach1,Samsonov_mach2,Havnes_mach3,Mamun_mach4,Hou_mach5}) are generated when a dust particle moves through a stationary dusty plasma medium. The topic of forewake excitations (precursor waves) has however received very limited attention till recently.  The first experimental demonstration of the generation of a precursor soliton in a flowing dusty plasma was reported in 2016 by Jaiswal \textit{et al}. \cite{surbhiprecursor}. In their study they observed the spontaneous excitation of precursor solitons when a super-sonically (with respect to the dust acoustic speed) moving dusty plasma fluid was made to flow over an electrostatic potential structure. The physical mechanism underlying this phenomenon can be understood in terms of the fluid concept used to explain hydrodynamic precursors. An object moving in a fluid always creates a pileup of matter in front of it. If its speed is sub-critical then the matter in front can disperse away at the linear phase velocity leading to the creation of a wake structure. However if the object speed is super-critical then the matter in front cannot disperse away fast enough and continues to build up. At a certain stage nonlinear effects become important and this can lead to the formation of solitons or other nonlinear structures. These nonlinear structures can move at a higher speed and can therefore separate from the object and move away as precursors. These first ever observations of precursor solitons in a plasma have been well characterized by detailed experimental measurements of their propagation features and qualitative comparisons with results from theoretical models. The existence of such precursor excitations have also been confirmed from full scale fluid simulations \cite{sanat_pinned}as well as molecular dynamic simulations \cite{forewake_sanat} . \par   
The object of our present study is to further consolidate and extend the previous experimental study \cite{surbhiprecursor} by carrying out a detailed investigation of the propagation characteristics of the precursors under varying excitation conditions. In particular, we look at the effects of varying the size and shape of the potential hill on the fore-wake phenomenon. We find that there is a systematic dependence of the amplitude, velocity and number of excited precursors on the size of the potential hill and the existence of a threshold value of the potential height below which no precursors are excited. There is also a very interesting dependence on the shape of the potential hill that highlights the role of sharp gradients in the excitation of precursors.\par
The paper is organized as follows: In the next section (in Sec. \ref{sec:setup}) we describe the experimental setup and procedure for excitation of precursor solitons and wakes. The experimental results of precursor solitons and wakes are discussed in Sec. \ref{subsec:precursor}. The excitation of these forced solitary waves over different sizes of the potential profile of the charged object are described in Sec. \ref{subsec:size} and the results on the influence of the shape of the potential profile are described in \ref{subsec:shape}. A brief summary of all the results and some concluding remarks are given in Sec. \ref{sec:conclusion}.      
\section{Experimental set-up and procedure}\label{sec:setup}
\begin{figure}[ht]
\includegraphics[scale=0.55]{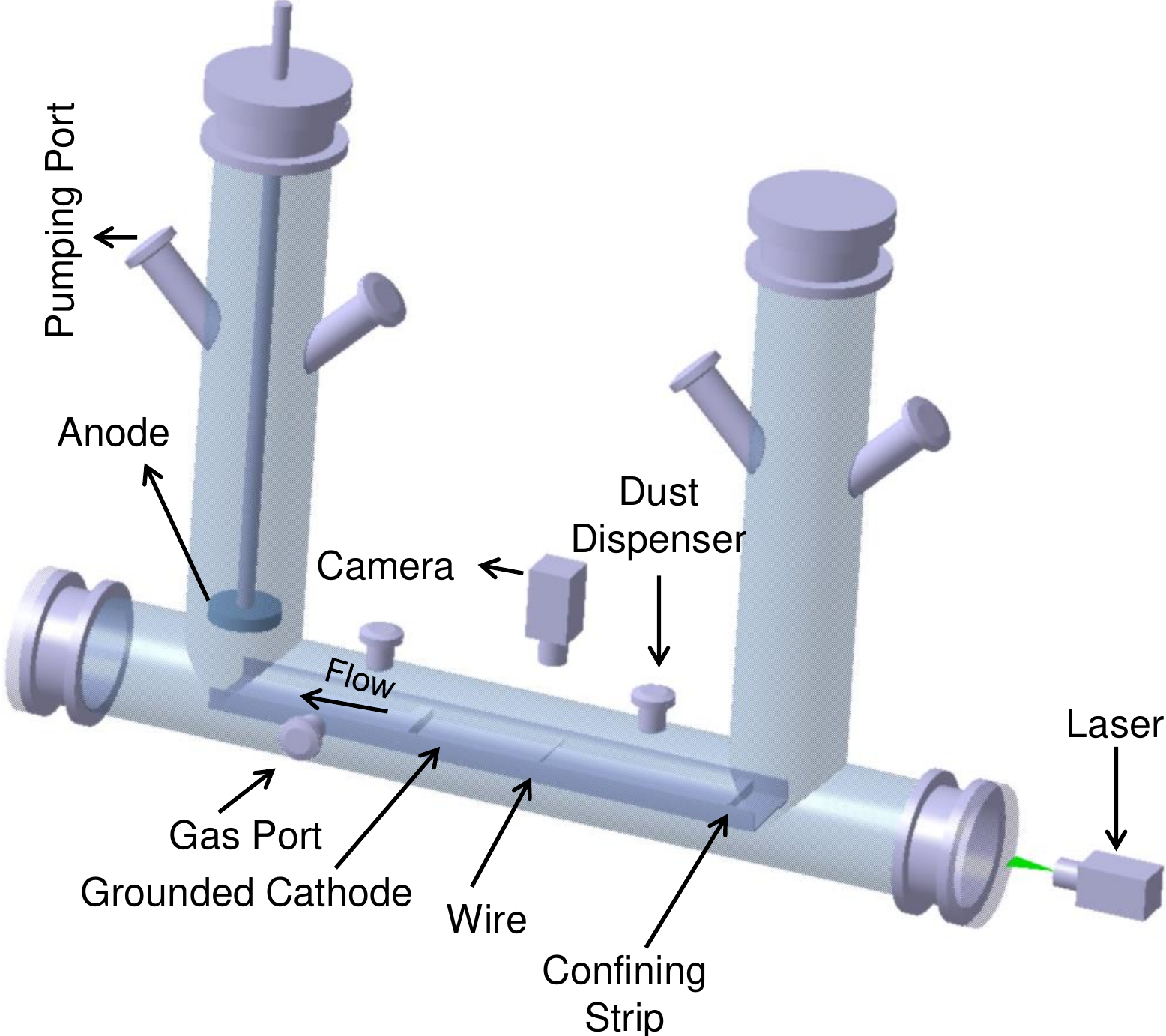}
\caption{\label{fig:fig1} A schematic diagram of dusty plasma experimental (DPEx) setup.  }
\end{figure}
The experiments are performed in a $\Pi$ shaped Dusty Plasma Experimental (DPEx) device and its schematic diagram is shown in Fig.~\ref{fig:fig1}. A detailed description of the experimental setup and its associated diagnostics are available in Ref. \cite{surbhirsi}.  DPEx device is basically an inverted $\Pi$ shaped glass chamber which has a number of axial and radial ports used for different purposes. A rotary pump is connected to one of the arms of DPEx device for evacuating the vacuum vessel whereas a mass flow controller is attached at the gas port to feed argon gas into the chamber.  A disc shaped anode and a tray shaped cathode are used for plasma generation. The tray type configuration of cathode is chosen for providing radial confinement to the dust particles. In addition to that two metal strips are also placed at the ends on the cathode to confine the particles axially. {For the experiments to study the effect of the obstacle height on the solitons}, a copper wire of diameter 1 mm and length 50 mm is mounted at a height of 10 mm from the base of the cathode. The wire acts as a charged object for the excitation of solitary waves as reported in Ref. \cite{surbhiprecursor}. This  wire is connected to the grounded potential through a switch and it can also be kept at various potentials by connecting a variable resistance ranging from 10 K$\Omega-10$ M$\Omega$ in between the wire and the switch. Micron sized Kaolin particles (2-5 $\mu$m) are sprinkled on the cathode in between the wire and the right strip before closing the chamber for creating a dusty plasma. These dust particles get negatively charged in the presence of the background plasma and levitate in the cathode sheath region where the sheath electrostatic force acting on the particle balances the gravitational force. The mass of the levitated dust particles $m_d\sim$ $8.9\times 10^{-14}$ kg for a radius of $r=2$ $\mu$m. These levitated micro particles are then illuminated by a green laser light and their dynamics are captured by a fast CCD camera for further analysis.
\par
\begin{figure}[ht]
\includegraphics[scale=0.5]{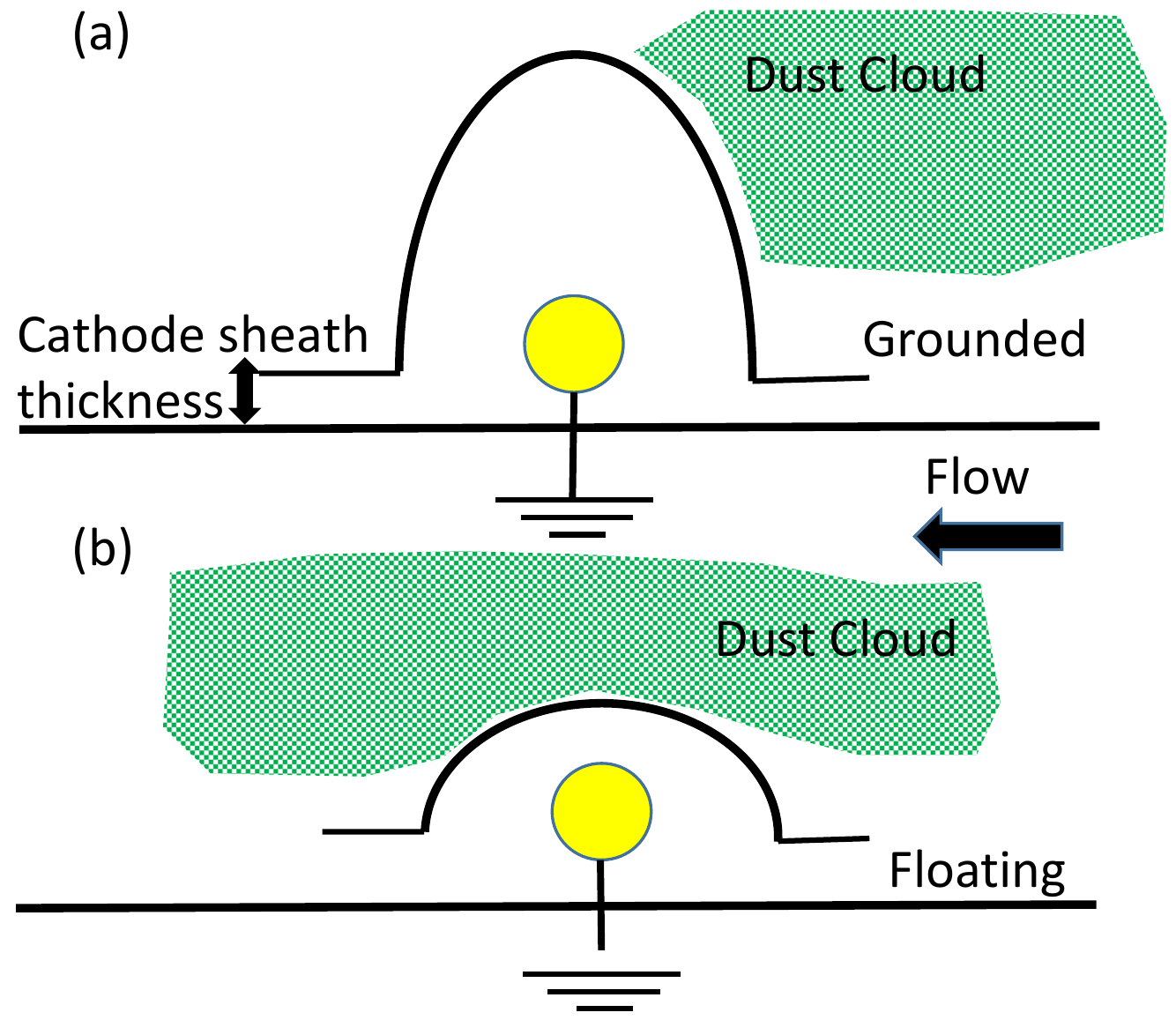}
\caption{\label{fig:fig2} (a) Equilibrium configuration of dust cloud before generating the flow. (b) Dust flow initiated by lowering the potential hill suddenly by making it floating. The yellow circle represents the location of charged object and flow is generated from right to left.}
\end{figure}

To start with, the chamber is pumped down to a base pressure 0.1 Pa and then Argon gas is injected into the device to set the  working pressure at 9-15 Pa. In this condition the gate valve attached with the pump is opened at 20$\%$ whereas the mass flow controller is opened at 5$\%$ to maintain the equilibrium pressure in the chamber. An Argon plasma is produced between the anode and grounded cathode by applying a voltage in the range of 280-320 V using a DC power supply. The plasma parameters like plasma density ($n_i$) and electron temperature ($T_e$) are measured using a single Langmuir probe and they come out to be $n_i\sim 0.5-3\times10^{15}$/m$^3$ and $T_e\sim 2-5$ eV for the range of discharge conditions as discussed in details in Ref. \cite{surbhirsi}. To create a dusty plasma, first the plasma is formed at a higher voltage and left for a few minutes. Later, the discharge voltage is reduced and set accordingly so that a highly dense dust cloud is formed. To repeat the earlier observations of Jaiswal \textit{et al.} \cite{surbhiprecursor}, the wire is kept at ground potential (negative compared to the plasma potential), which acts as a barrier for these negative charged dust particles and as a result the particles remain confined between the wire and the right potential strip as shown in Fig.~\ref{fig:fig2}(a) and exhibit only thermal motion. For the present set of experiments, the discharge condition is kept in such a way that there is no spontaneous excitation of waves in dust cloud due to ion streaming. For a specific discharge condition, $V_d= 320$ V and p = 11 Pa, the average charge of these micron sized dust particles comes out to be $Q_d\sim$ $8.43\times 10^3e$ which is estimated from a Collision Enhanced plasma Collection Model (CEC) \cite{khrapakcec,khrapakcec2}. The other dust parameters like dust density $n_d\sim 10^{11}/m^3$ and dust temperature $T_d\sim 0.6-1.2$ eV are estimated with the help of an IDL based super Particle Identification Tracking (sPIT) code \cite{konopka2000wechselwirkungen,fengrsi}. Based on the above parameters the phase velocity of the Dust Acoustic Wave (DAW) turns out to be $C_{da} \sim 20$ mm/sec. To compare the theoretical estimation, a separate experiment is also carried out to excite DAW for the above discharge condition by applying a sinusoidal signal to the wire.  It is found that the measured $C_{da}$ comes out to be $\sim 25$ mm/sec, which is in good agreement with the estimated value.  \par
For the excitation of forced solitary waves and wake structure, the dust particles are made to flow over the wire. The flow of the dust cloud is generated by suddenly lowering the height of the negative potential hill created by the charged object. When the wire is switched to near floating potential the potential hill is lowered and the dust particles flow from right to left over the wire as shown in Fig.~\ref{fig:fig2}(b). During the flow, the dust particles attain a constant velocity due to neutral drag force\cite{surabhi_psst,garima_second} and excite solitary waves along with wakes depending on the flow velocity of dust fluid. The flow velocity of dust fluid is measured by tracking the individual particles at the tail of the cloud using Particle Image Velocimetry (PIV) technique \cite{thielicke2014pivlab}.  The velocity of the dust fluid is changed from subsonic ($\sim 15-18 $~mm/sec) to supersonic ($\sim 25-32$~mm/sec) values by varying the discharge parameters. To study the propagation characteristics of the excited structures, the size (both height and width) of the potential hill created by the charged object is altered by changing the resistance connected to the wire, whereas the shape of the potential profile is modified by using objects of different shapes. These studies are discussed in the subsequent sections (Sec. \ref{subsec:size} and Sec. \ref{subsec:shape}).
\section{Results and Discussion}\label{sec:results}
\subsection{Excitation of precursor solitons and wakes}
\label{subsec:precursor}
\begin{figure}[hb]
\includegraphics[scale=0.75]{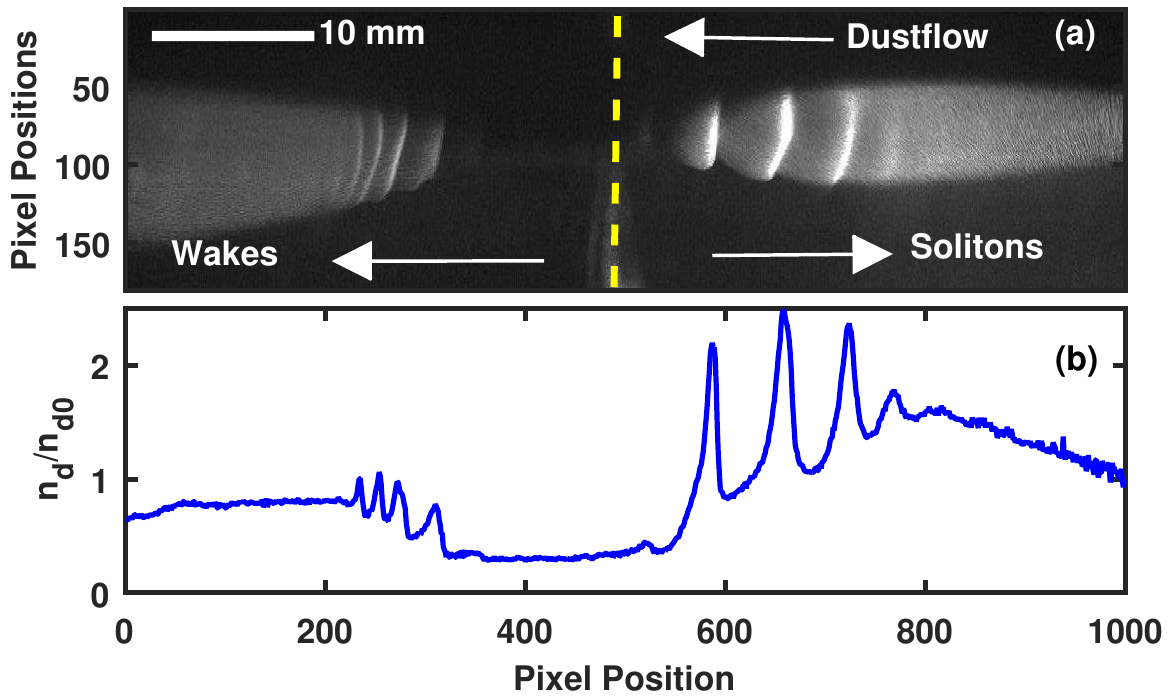}
\caption{\label{fig:fig3} (a) Propagation of solitons against the direction of flow (in upstream direction) and wakes in the direction of flow (in downstream direction) for the case of supersonic flow. The position of wire is represented by the yellow dashed line. (b) The intensity profile of the solitons and wakes.}
\end{figure}

Before investigating the dependence of shape and size of the potential hill on the propagation characteristics of precursor solitons, the experiments of {\color{blue} Jaiswal} \textit{et al.} \cite{surbhiprecursor} were repeated to serve as a benchmark. The results are briefly summarized here. As in \cite{surbhiprecursor},  for a subsonic flow only wake structures were observed in the downstream direction whereas supersonic flows created precursor solitons propagating in the upstream direction ahead of the object and wakes in the downstream direction. Fig.~\ref{fig:fig3}(a) depicts such a scenario for a supersonic flow of dust fluid, with Fig.~\ref{fig:fig3}(b) showing the normalized amplitudes of solitary and wake structures extracted from Fig.~\ref{fig:fig3}(a). \par
\begin{figure}
\includegraphics[scale=0.72]{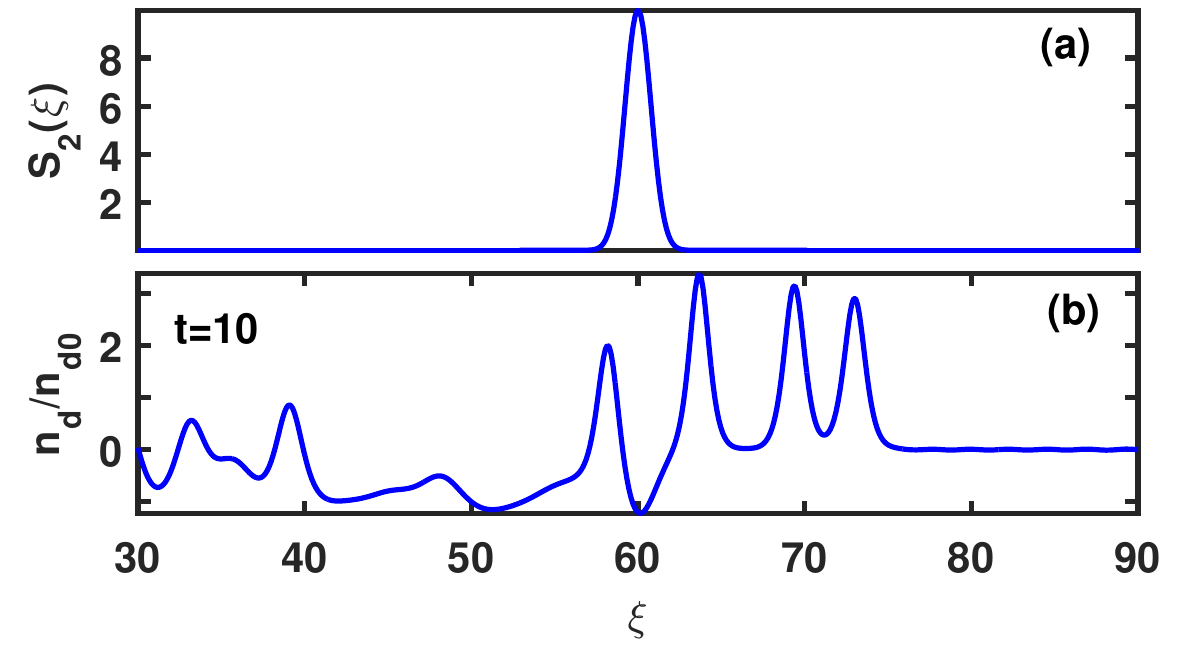}
\caption{\label{fig:fig4} (a) Source function (b) Precursor solitons and wakes obtained from the numerical solution of f-KdV equation.}
\end{figure}

These nonlinear precursor structures were confirmed to be solitons by checking the constancy of the  quantity amplitude $\times$ width$^2$. The following model f-KdV \cite{SEN2015429} equation was also solved as part of the benchmark exercise to obtain a qualitative account of the phenomenon.
\begin{eqnarray}
\frac{\partial n_{d1}}{\partial t}+A\frac{\partial n_{d1}}{\partial \xi}+\frac{1}{2}\frac{\partial^3 n_{d1}}{\partial \xi^3}=\frac{1}{2}\frac{\partial S_2}{\partial \xi},
\label{eq:eq1}
\end{eqnarray}
where the source term ($S_2$) represents the charged object, $n_{d1}$ is the perturbed density normalized to the equilibrium density $n_{d0}$ and $\xi(t)$ is the coordinate in the wave frame moving at the phase velocity $u_{ph}$ normalized to the dust acoustic speed. The coefficient \lq A' associated with Eq.~\ref{eq:eq1} depends upon the electron and ion temperatures and the plasma density. A detailed description of the f-KdV model equation and its solution can be found in \cite{SEN2015429}. The source function $S_2$, as shown in Fig.~\ref{fig:fig4}(a), is chosen as a Gaussian with amplitude $A_s$ and width $G$. Fig.~\ref{fig:fig4}(b) shows the spatial variation of density perturbation which is obtained by solving f-KdV equation (Eq.~\ref{eq:eq1}) numerically. In this simulation, the Gaussian source function is made to move from left to right either subsonically or supersonically to replicate the relative velocity between the wire and the dust flow in our experiments and the coefficient $A$ is chosen to be 4 for our experimental parameters.  As shown in Fig.~\ref{fig:fig4}(b), just like the experimental observation, the numerical solution of f-KdV equation yields the excitation of precursor solitons which propagate in the upstream direction whereas the wakes move in the downstream direction. The f-Kdv will be further used to model the present set of experiments exploring the effect of the size and shape of the obstacle on the solitonic excitations.
\subsection{Excitation of precursor solitons for flow over objects of different sizes} 
\label{subsec:size}
The benchmark experiments were executed by changing the height of the potential hill from its maximum value (grounded potential) to its minimum value (floating potential) to excite the precursor solitons. To  investigate the effect of the size (height and width) of the potential hill on the phenomenon,  a systematic set of experiments were performed by connecting a variable resistance in series with the wire to change the size of the potential hill. The size of the potential hill of the charged object is adjusted by drawing a current through various resistance values ranging from 10 K$\Omega$ to 10 M$\Omega$ and measuring the voltage across the resistance ($V_{wg}$).  
\begin{figure}[ht]
\includegraphics[scale=0.49]{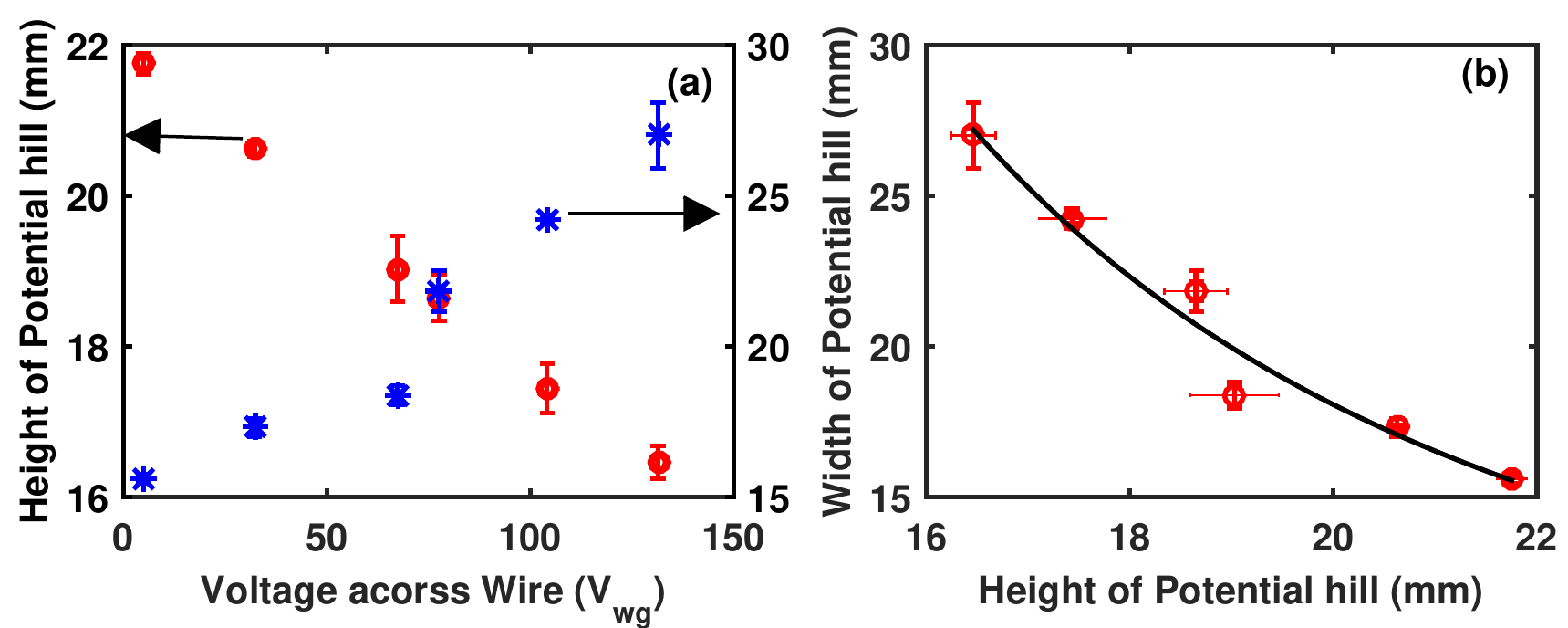}
\caption{\label{fig:fig5} (a) Variation of height (represented by red open circles) and width (represented by blue star) of the potential hill with $V_{wg}$. (b) Power law fit on the height and width of the  electrostatic potential created by the charged wire.}
\end{figure}
To begin with, the height and width of the potential hill is measured using dust particles as a dynamic micro probes as discussed in Ref.~\cite{garima_first}. In this technique, these tracer particles are made to flow over the potential hill and tracking of the individual particle trajectories yields their positions and velocities over the time. The potential of the electrostatic hill just above the wire is then directly estimated by using the energy conservation arguments. The height and the width of the potential hill is measured for a wide range of $V_{wg}$ and plotted in Fig.~\ref{fig:fig5}. Fig.~\ref{fig:fig5}(a) shows the variation of height (represented by \lq o') and width (represented by \lq$\star$') of the potential hill created by the charged object with voltage measured ($V_{wg}$) across the different resistance connected with the wire in series. It can be seen that the height of the hill decreases and width increases with the increase of $V_{wg}$. The height of the potential hill is maximum when $V_{wg}$ becomes close to the grounded potential which is negative with respect to the plasma potential and it decreases when the voltage drop across the variable resistance reaches towards the plasma potential which is $\sim 300$ V for this specific set of experiments. The variation of the height (h) with width (w) of the potential hill created by the wire is plotted in Fig.~\ref{fig:fig5}(b) and it is seen that the width decreases with an increase of the height of the potential hill. The solid line represents the power law relationship ($h\sim w^{-3.13}$) between the width and height. Further experiments are performed with similar range of $V_{wg}$ to examine the effect of the size of the potential hill on the excitation of forced solitary waves.\par
\begin{figure}[ht]
\includegraphics[scale=0.75]{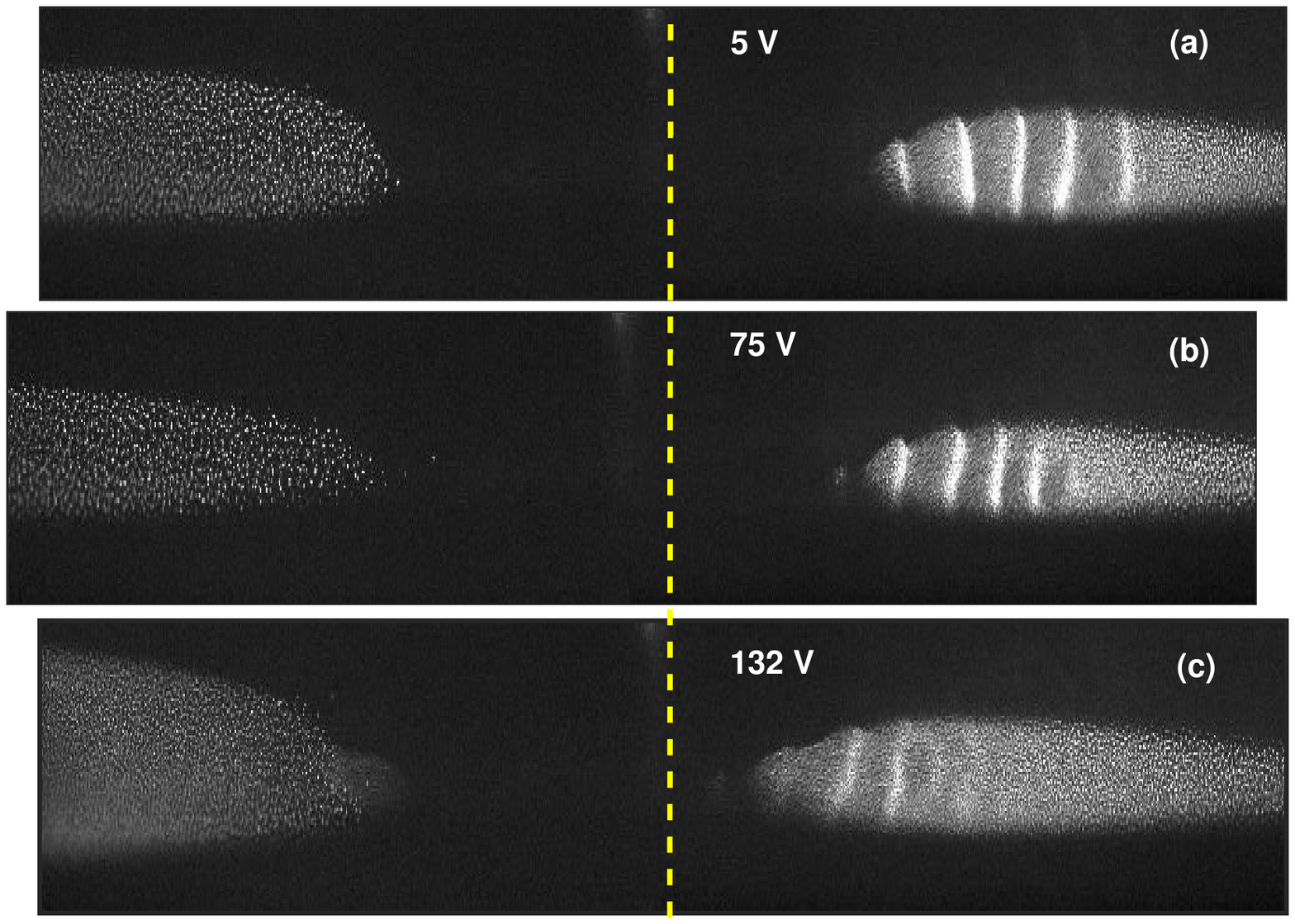}
\caption{\label{fig:fig6} Excitation of precursor solitons for (a) 5 V, (b) 75 V, (c) 132 V by varying the voltage across the hill created by the charged wire.}
\end{figure}
Figure~6(a-c) shows typical images of fully developed forced solitary structures. It visually shows how the number, amplitude and width of solitons changes with the wire potential, $V_{wg}$. Fig.~\ref{fig:fig6}(a) shows the solitary waves at $p=$ 11 Pa and $V=$ 320 V when $V_{wg}$ is kept at 5 V which is approximately at grounded potential. The dotted line in this figure represents the location of the wire. For $V_{wg}=5$ V, five prominent solitary waves get excited which are found to propagate in the upstream direction from the frame of fluid. When the resistance connected to the wire is increased such that $V_{wg} \sim 75$ V, the number of prominent  solitary structures decreases to four as shown in  Fig.~\ref{fig:fig6}(b). The number further decreases to two when the $V_{wg}$ is increased to 132 V as shown in Fig.~\ref{fig:fig6}(c). If the potential of the wire is increased further towards the floating potential, the height of the source object decreases and it is not sufficient  to excite any forced solitary structures. We have also found the threshold value, $V_{wg}=135$ V, beyond which the particles simply flow and do not excite any wave structures.  It is also found that the amplitude decreases whereas the width of the fully developed solitons increases with the increase in $V_{wg}$ but the change is not visually appreciable. A more quantitative discussion of the variation of amplitude, width and velocity with  $V_{wg}$ is given later in this section. In all these experiments ({Fig.~\ref{fig:fig6}(a-c)), the dust velocity is measured using PIV analysis to make sure that the dust fluid moves super-sonically over the wire. \par
\begin{figure}[ht]
\includegraphics[scale=0.48]{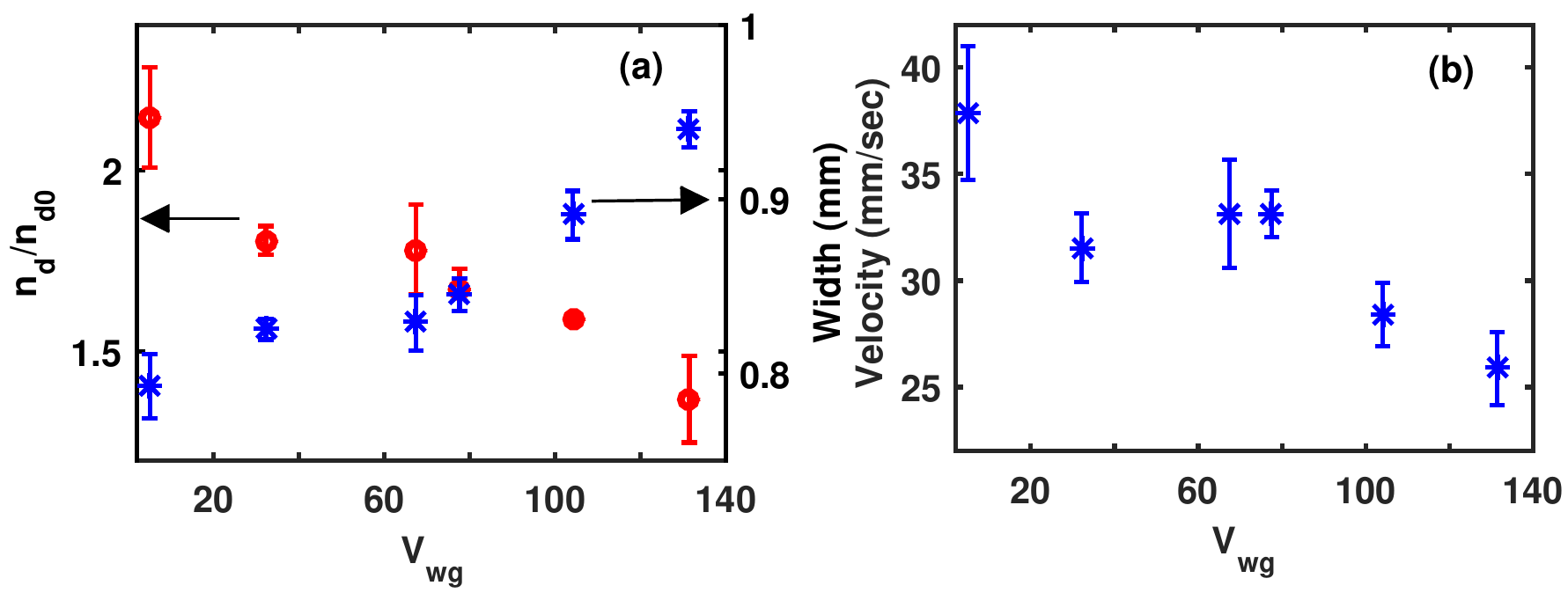}
\caption{\label{fig:fig7} Variation of (a) amplitude (represented by red open circles), width (represented by blue star) and (b) velocity of the excited precursor solitons for the case of supersonic fluid flow with the voltage drop $V_{wg}$ across the resistance. }
\end{figure}
\begin{figure}
\includegraphics[scale=0.75]{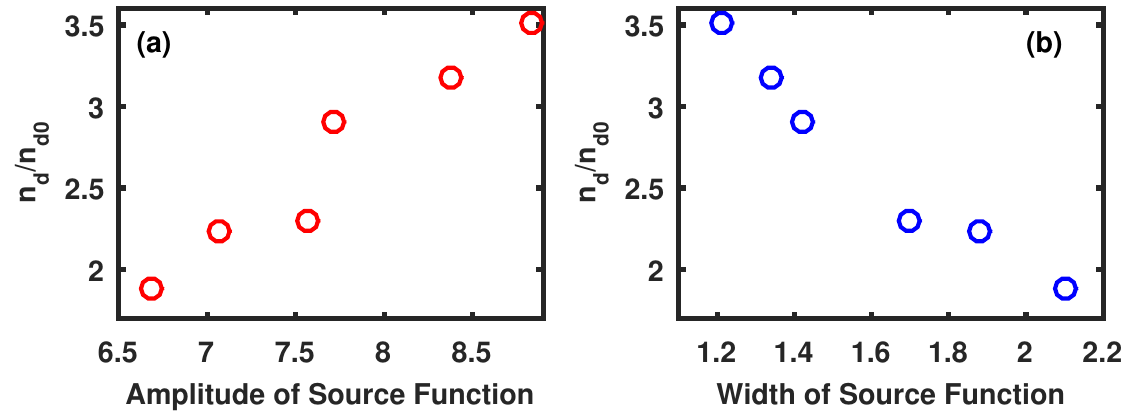}
\caption{\label{fig:fig8} Variation of amplitude of solitons with (a) amplitude of source, (b) width of source numerically obtained from the solution of f-KdV.}
\end{figure}
Figure~\ref{fig:fig6} qualitatively shows that the amplitude  of the solitary structures decreases whereas width increases with the increase of $V_{wg}$. For a quantitative analysis of the amplitude ($n_d/n_{d0}$) of solitary wave, the maximum intensities are extracted from the images later normalized by the equilibrium intensity of dust density. Fig.~\ref{fig:fig7}(a) shows the variation of this amplitude of forced dust acoustic solitary waves (f-DASw) with $V_{wg}$. It shows that the amplitude of these excited solitary waves decreases whereas the width increases with the increase in $V_{wg}$. It can also be seen in Fig.~\ref{fig:fig5}(a), that the height of the potential hill decreases whereas the width increases with the increase of $V_{wg}$. It essentially indicates that the amplitude of the solitary structures strongly depends on the size of the potential hill. In addition to that the velocity of the forced dust acoustic wave  also decreases with the increase of $V_{wg}$ as shown in Fig.~\ref{fig:fig7}(b).  Hence in our experiments, the solitary wave having higher amplitude propagates with higher velocity which is one of the important properties of a KdV type solitons. Therefore, it can be concluded that the size of the potential hill plays a salient role on the propagation characteristics of a solitary wave. \par
To model these experimental findings, we have again solved numerically the f-KdV equation (Eq. I) for different values of the amplitude ($A_s$) and width ($G$) of the Gaussian source function by choosing the values from the power law relationship derived from the experiments and measured the mean value of the amplitude of solitary structures. Fig.~\ref{fig:fig8} depicts the variation of amplitude of forced solitary waves obtained from numerical solution of f-KdV with the height and width of the source function. For the present set of numerical simulation, the amplitudes and widths are chosen from the experiments as shown in Fig.~\ref{fig:fig5}(b). In Fig.~\ref{fig:fig8}(a) one can see that the amplitude of the excited forced solitary waves decreases with the decrease of amplitude of the source function while it decreases with the increase of width of the source function as shown in Fig.~\ref{fig:fig8}(b). The first point of Fig.~\ref{fig:fig8}(a) and last point of Fig.~\ref{fig:fig8}(b) correspond to a single soliton solution of the equation which is also the threshold conditions for the creation of f-KdV solitons. Further decreasing of A and increasing of G causes only the excitation of wake structures instead of the creation of solitons.} Therefore, the findings from the numerical analysis qualitatively support our experimental observations and there is indeed a threshold for the generation of solitons in the f-KdV simulation similar to the experiments.
\subsection{Forced Solitary Waves over different shapes of the charged object}
\label{subsec:shape}
\begin{figure}[ht]
\includegraphics[scale=0.35]{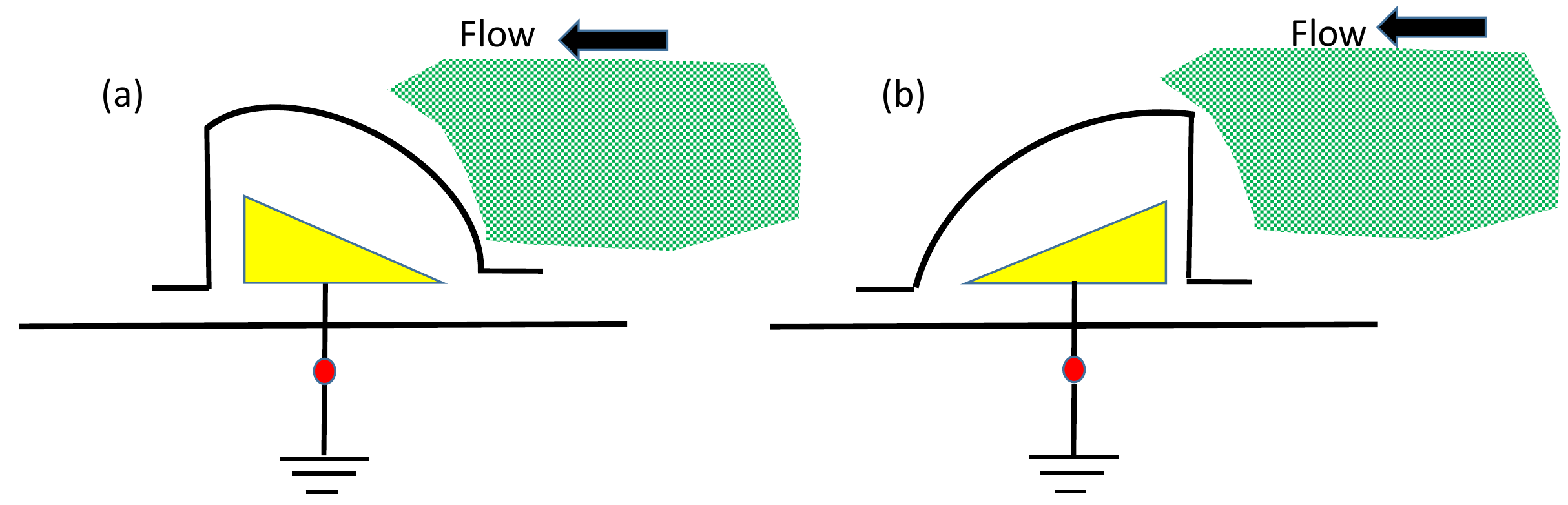}
\caption{\label{fig:fig9}Equilibrium configuration of dust cloud before generating the flow for two different shape of the charged object.}
\end{figure}

All the above experiments were performed by flowing the dust fluid supersonically over the wire of different sizes (height and width) of the potential hill. The potential profile around the wire \cite{garima_first} is indeed a Gaussian where the fall of the sheath profile is symmetric and sharp around the object. Next, to investigate the effect of shape of potential profile on the propagation characteristics of forced solitary waves, a triangular shaped object was used in two different configurations as shown in Fig.~\ref{fig:fig9}(a) and (b). The supersonic flow in the dust fluid was again generated by switching the object from grounded potential to floating potential as described in Fig.~\ref{fig:fig2}. In the first case (see, Fig.~\ref{fig:fig9}(a)), the dust fluid flows over the rising slope of the potential profile created by the triangular charged object, whereas in another configuration, dust fluid flows over a steep charged object  (see, Fig.~\ref{fig:fig9}(b)) first and then flows over falling slope.\par  
\begin{figure}[ht]
\includegraphics[scale=0.70]{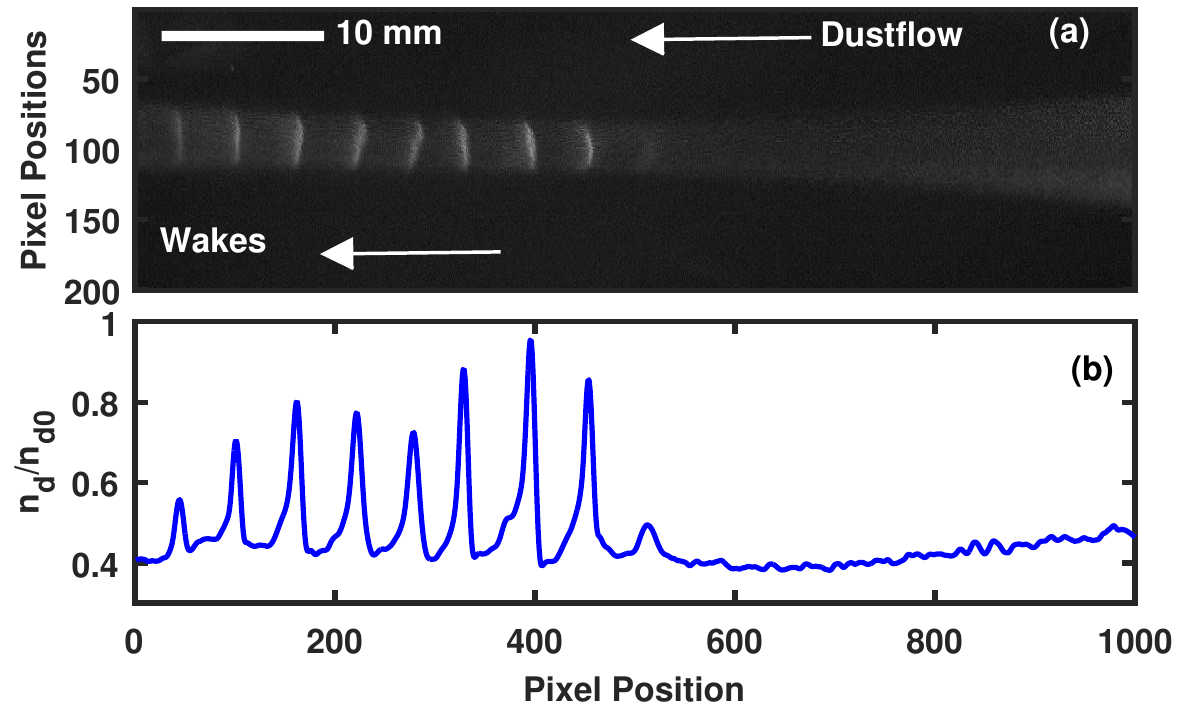}
\caption{\label{fig:fig10} (a) Generation of non linear wakes in the direction of flow when the flow is initiated and facing rising slope of the potential hill (b) Intensity profile of Fig.~\ref{fig:fig10}(a).}
\end{figure}

Figure~\ref{fig:fig10}(a) shows a typical image of the excited structures when the dust fluid supersonically flows over the charged object as shown in Fig.~\ref{fig:fig9}(a). Interestingly in this configuration, no structures are found to propagate in the upstream direction dissimilar to the case when the object is purely Gaussian (see Fig.~\ref{fig:fig3}(a)). Instead a series of crests are found to propagate in the down stream direction. The perturbed dust density profiles (extracted from Fig. \ref{fig:fig10}(a)) is plotted in Fig.~\ref{fig:fig10}(b) to explore the properties of these structures. These smaller amplitude crests propagate in the downstream direction with respect to the frame of fluid as wakes. 
\begin{figure}[ht]
\includegraphics[scale=0.65]{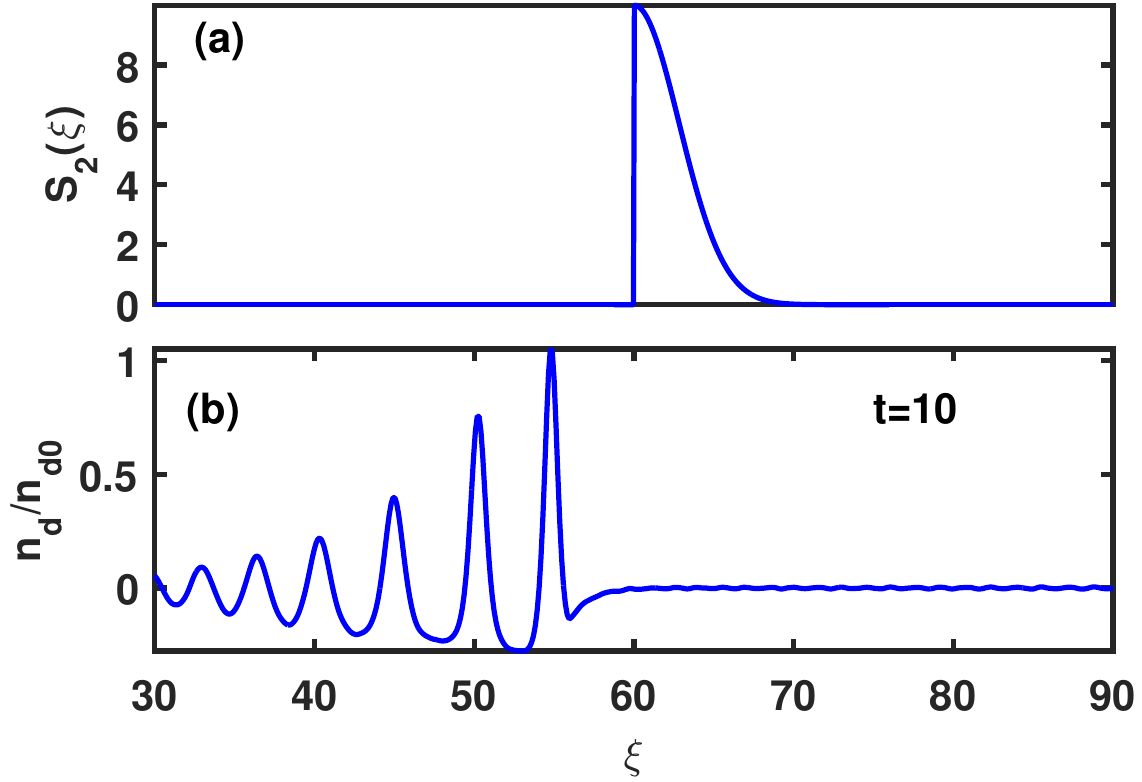}
\caption{\label{fig:fig11} Time evolution of numerical solution of f-KdV equation using the source function half Gaussian function as Source function to replicate the triangular shaped object.}
\end{figure}
These experimental findings are also compared with numerical solutions of forced-KdV equation (see Eq.~I) in which the source function is taken as a half Gaussian with a rising slope as shown in Fig.~\ref{fig:fig11}(a).  In case of the triangular object with flat hypotenuse (Fig.~\ref{fig:fig11}(a) and Fig.~\ref{fig:fig13}(a)), the object is replicated by a half Gaussian with a width ($\frac{W}{2} \sim 6$) that is bigger than the source function with a full Gaussian. The choice of the width of the source term is dictated by the size of the sheath that forms around the obstacle. For p=11 Pa and V= 320 V the sheath around the grounded wire of 1~mm diameter is $\sim 15.5$~mm (see first \lq$\star$' of Fig.~\ref{fig:fig5} (a)). For the same discharge condition, the sheath around the triangular charged object (of hypotenuse 65 ~mm) is $\sim$ 80.5~mm. Hence, $\frac{W}{2}$ for the half Gaussian source function is chosen to be 6 which is approximately 5 times higher compared to the full Gaussian. Fig.~\ref{fig:fig11}(b) displays the perturbed dust density profile for a given time when the object moves with supersonic velocity similar to our experimental situation. The solution of f-KdV equation shows only the excitation of wake structures, which propagate in the downstream direction as observed in the experiments. \par
\begin{figure}[ht]
\includegraphics[scale=0.70]{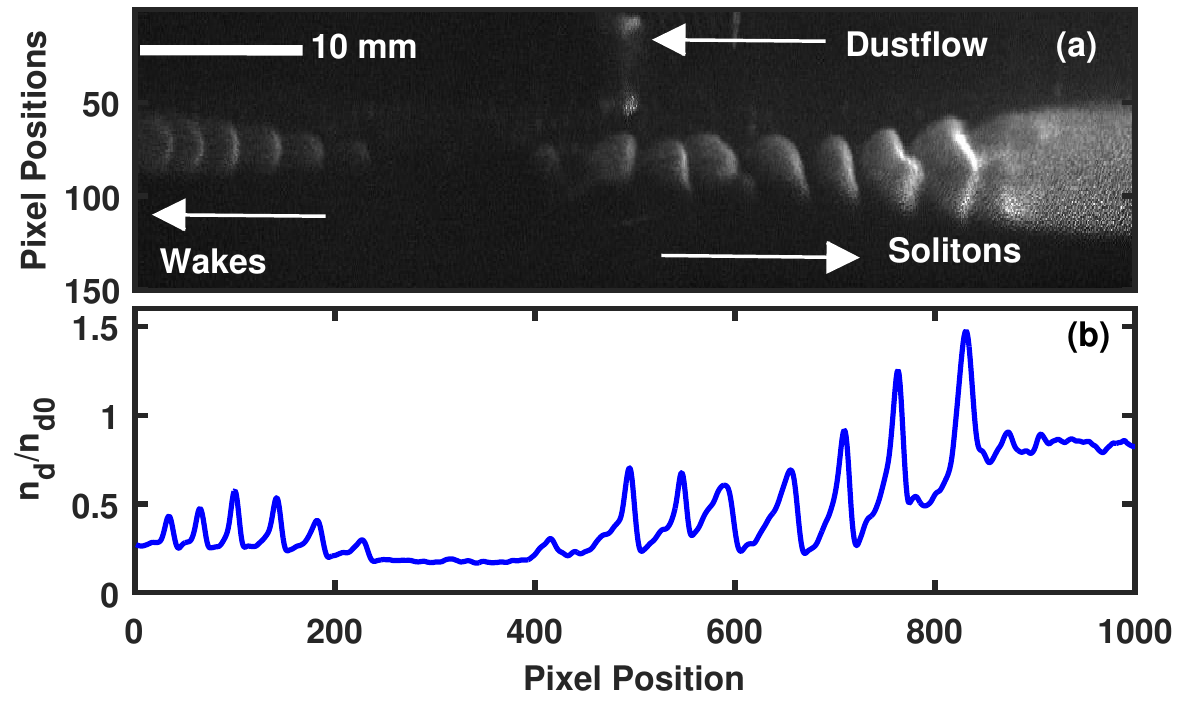}
\caption{\label{fig:fig12}(a) Generation of wakes and solitons when the flow faces sharp rise of the potential hill (b) intensity profiles of these wakes and solitons extracted from fig.~\ref{fig:fig12}(a).}
\end{figure}
\begin{figure}[ht]
\includegraphics[scale=0.70]{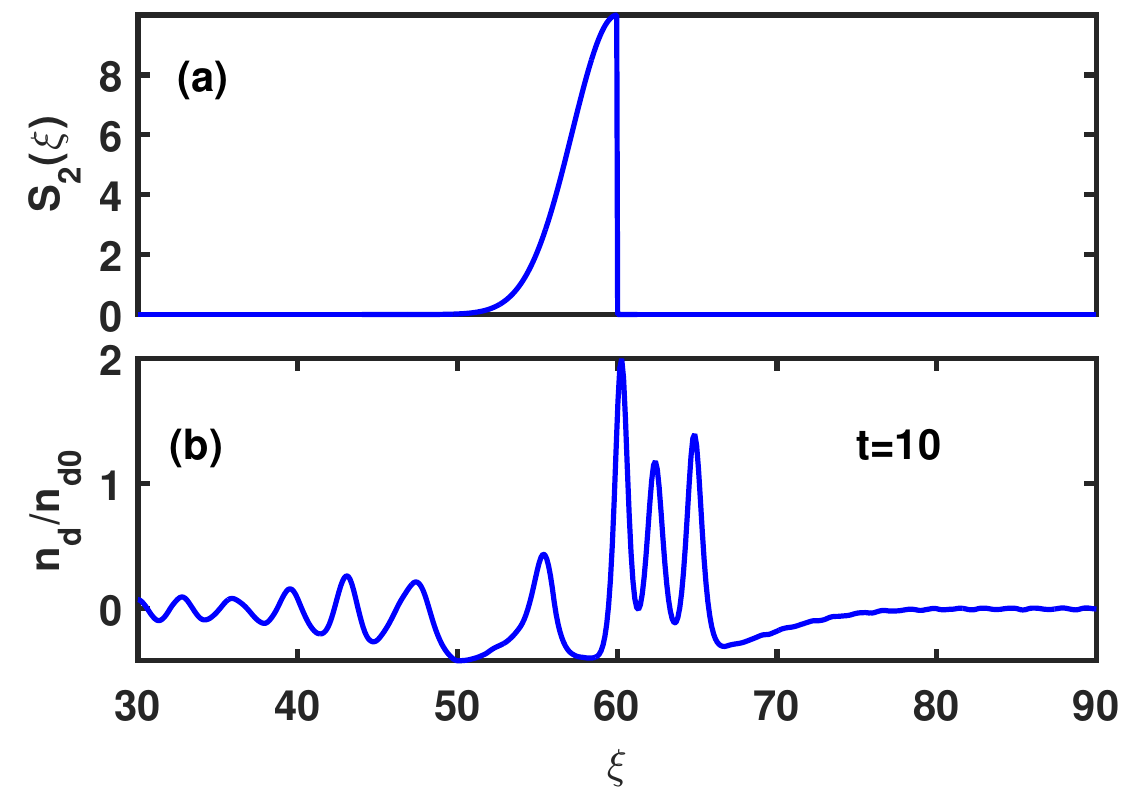}
\caption{\label{fig:fig13} (a) Source function (b) Time evolution of numerical solution of f-KdV equation by reversing the source function as shown in fig.~\ref{fig:fig11}(a).}
\end{figure}

To investigate further the influence of shapes, the same triangular object is used in reverse configuration as shown in Fig.~\ref{fig:fig9}(b). In this configuration, the dust fluid first faces a sharp rise of the potential and then it goes through a monotonic fall in the potential.  When the fluid flows supersonically over this object, it excites both precursor solitons in the upstream direction as well as wakes in the downstream direction as shown in Fig.~\ref{fig:fig12}(a). The perturbed dust density extracted from Fig.~\ref{fig:fig12}(a) is shown in Fig.~\ref{fig:fig12}(b). 
To mimic the experimental observations, the f-KdV equation is again solved numerically by using a half Gaussian object like Fig.~\ref{fig:fig11}(a) but in the reverse direction as shown in Fig.~\ref{fig:fig13}(a). The solution of f-KdV also shows the excitation of solitary waves in the upstream direction and wakes in the downstream direction. Hence, it can be concluded  further that the shape of the charged object is  also an important parameter which decides the excitation of nonlinear structures. A linear gradient is insufficient to excite the precursor solitons and only gives rise to the customary wakes whereas a sharp gradient excites both.         
\section{Summary and Conclusion}
\label{sec:conclusion}
A set of experiments were done to investigate in detail the modifications of the propagation characteristics of precursor solitons which get excited when a supersonic flow of dust fluid is made to flow over a stationary charged object of different sizes and shapes. The experiments were performed in the DPEx device in which dusty plasma was created in a DC glow discharge Ar plasma using Kaoline particles. The flow of the dust cloud was initiated by suddenly lowering the height of the potential hill created by a charged object (wire or metallic wedge shaped object). For the case of subsonic dust fluid flow, only wakes were observed in the downstream direction in the frame of the fluid whereas for a supersonic flow solitary structures were excited in the upstream direction along with the wakes in the downstream direction. The size and shape of the potential profile was varied by connecting a variable resistance in series with a copper wire and by using the triangular object in different orientation with respect to the flow, respectively. The major  outcomes of our present work are listed below: 
\begin{itemize}

\item To investigate the effect of size of the potential hill on the propagation of precursor solitons, the height of the potential hill was varied by drawing current through a variable resistance. It was found that the amplitude and velocity of the solitons decreased whereas their widths increased with a decrease in the height of potential hill. The number of emitted solitonic structures also decreased with the decrease of the hill height. It was also discovered that there is a threshold height of the potential hill below which the dust fluid simply flows over the object without exciting any structures.  

\item The f-KdV equation was once again solved numerically by changing the height and width of the Gaussian source function. The numerically obtained amplitudes and widths of solitonic structures followed a similar trend as observed in the experiments.

\item In another set of experiments, the shape of the potential profiles of the charged object was changed by replacing the wire by a solid triangular shaped object. In this specific set of experiments, the flow was only generated by switching  the charged object from ground to floating potential. When the dust fluid flowed supersonically over this triangular object facing the linearly increasing slope only wakes were found to be excited in the downstream direction and no nonlinear structures were seen to propagate in the upstream direction. However, when the experiments were carried out by reversing the object so that the dust fluid faced the sharp jump of the potential created by the charged object, both solitons propagating in the upstream direction and wakes propagating in the downstream direction were seen.
\item These results were also modeled and compared qualitatively with the experimental findings by taking the source function to be a half Gaussian in the forward and reverse directions respectively.
\end{itemize}
Hence to summarize, the excitation of precursor solitons and their propagation characteristics are found to depend on the shape and size of the potential profiles of the charged object over which the fluid flows. Our experimental  findings not only confirm and consolidate earlier experimental observations of precursor solitons but bring to light new fundamental results regarding the excitation process. These insights should prove helpful in interpreting the manifestation of this concept in natural occurrences such as in excitations triggered by solar wind interactions with the earth and moon or space debris or satellite interactions with the ionospheric plasma.  The findings may also stimulate further experimental and theoretical studies towards more fundamental investigations of this yet poorly explored topic in plasma physics. \\\\\noindent
\textbf{ACKNOWLEDGMENTS}\\\\
A.S. is thankful to the Indian National Science Academy (INSA) for their support under the INSA Senior Scientist Fellowship scheme.\\\\
\textbf{References}        
%

\end{document}